\documentclass[aps,prl,twocolumn,groupedaddress]{revtex4}

\usepackage{graphicx}
\usepackage{amsmath}

\begin{document}
\title{Gap formation and soft phonon mode in the Holstein model}


\author{D. Meyer}
\email[]{d.meyer@ic.ac.uk}
\author{A. C. Hewson}
\affiliation{Department of Mathematics, Imperial College, London, SW7 2BZ, UK}

\author{R. Bulla}
\affiliation{Theoretische Physik III, Elektronische Korrelationen und Magnetismus, Universit{\"a}t Augsburg, 86135 Augsburg, Germany}

\date{\today}

\begin{abstract}
We investigate electron-phonon coupling in many-electron systems using
dynamical mean-field theory in combination with the numerical renormalization
group. This non-perturbative method reveals significant precursor effects to the
gap formation at intermediate coupling strengths. The emergence of a soft phonon
mode and very strong lattice fluctuations can be understood in terms of
Kondo-like physics due to the development of a double-well structure
in the effective potential for the ions.
\end{abstract}

\pacs{}

\maketitle

Despite the many years of study of the electron-phonon interaction in metallic
systems, there remain fundamental problems that have yet to be resolved;
particularly in the strong-coupling regime and in conjunction with strong
electron-electron interactions.
A solution to these problems will be required to understand fully phenomena
such as the colossal magnetoresistance effect in
manganites~\cite{MLS95}.
Also in the metallic alkaline-doped C$_{60}$-based compounds, 
high critical superconducting temperatures have been observed~\cite{Gun97}.
These materials are known to have strong
electron-phonon and electron-electron interactions~\cite{Gun97}.
Recent photoemission experiments indicate strong 
electron-phonon coupling in the cuprate high-temperature
superconductors~\cite{Lea01}.
There is a clear need of theoretical techniques to tackle these problems in the
strong-coupling regime. 

Although electron-phonon problems involving one or few electrons can be solved
to very high accuracy~\cite{TBK01,CPFF97}, so far there are no comparably accurate approaches
for the many-electron case relevant to metallic systems.
In this letter we study the simplest realization of
electron-phonon coupling: The Holstein model with finite
electron density describes the coupling of Einstein (LO) phonons
to the density of electrons of a non-degenerate conduction band:
\begin{equation}
  \label{eq:hamil}
  H=\sum_{\vec{k}\sigma} \epsilon(\vec{k}) c_{\vec{k}\sigma}^\dagger
  c_{\vec{k}\sigma} +
  \sum_i \omega_0 b_i^\dagger b_i + \sum_i g (b_i^\dagger +b_i) \sum_{\sigma} 
  (n_{i\sigma} -\frac{1}{2}).
\end{equation}
No general exact solution of this model is known
for systems with finite electron density, even in the limit of infinite spatial
dimensions ($d=\infty$). This limit takes local quantum fluctuations fully into
account and has proved to be a powerful tool in understanding
strongly correlated systems~\cite{MV89,GKKR96}.
Although exactly solvable for $d=\infty$,
the case of a single electron in the band~\cite{CPFF97} is physically very
different from the many-electron case since no electron-electron pairing
(bipolaron formation, superconductivity etc.) can occur.
Another, more instructive limiting case is
the static limit $\omega_0=0$, where
the phonons are replaced by a static displacement of the
lattice ('static' or 'classical' approximation). Extensive calculations in
this limit for $d=\infty$ have been presented in Ref.~\onlinecite{MMS96a}. 
However, it
is immediately clear that this static limit neglects all possible effects
stemming from the quantum nature of the lattice excitations.
In the opposite limit of $\omega_0\rightarrow\infty$
the lattice reacts instantaneously to the
state of the electrons.
Here, the Holstein model can be mapped
onto a non-retarded attractive Hubbard model~\cite{FJ94} by integrating out the
phonons. The Hubbard model has been intensively studied, and much recent
progress has been based on using the $d=\infty$ limit.
Although the large-$\omega_0$ limit is not physically relevant, it is still
a useful point of reference for getting an
overall understanding of the physics of the model.
Of physical concern for applications
are relatively small phonon frequencies of the order of $\omega_0
\approx 0.01 W - 0.2W$ ($W$ is the width of the
electron band). In this regime, the Migdal-Eliashberg diagrammatic
approach has been used~\cite{HD01pre}.
The main feature of this approach is the neglect of vertex corrections.
A sufficient criterion for its application is usually
$\omega_0/W \ll 1$. However, at least for half-filling
there is evidence that this approach breaks down for intermediate coupling
strengths $g$~\cite{HD01pre,AKR94}. There have also been a number of perturbative schemes
going beyond the Migdal-Eliashberg approach and including some vertex
corrections~\cite{BM01,DM02a,FJ94,BZ98}.

In the \textit{dynamical mean-field theory} (DMFT) a lattice-model is mapped
onto an associated impurity model. 
The parameters of the associated impurity model are related to the Green's
function of the lattice model by a \textit{self-consistency condition}.
This mapping becomes exact in the limit of infinite spatial dimensions ($d=\infty$).
The method is described in detail in Ref.~\onlinecite{GKKR96}. 
One of the most
precise techniques for solving the associated impurity model for low temperatures is
the \textit{numerical renormalization group} (NRG)\cite{KWW80a,CHZ94,SSK89}. 
It is capable of
resolving very low energy scales, and gives information about the excitation
spectrum over the whole real-energy axis.
The combination of DMFT and NRG has helped to solve a number of open questions
regarding the Mott transition in the Hubbard model~\cite{GKKR96,Bul99,BCV01}.
The application of the DMFT to the Holstein model~(\ref{eq:hamil}) leads to the
Anderson-Holstein impurity model, which is essentially an Anderson model with
additional coupling of a local phonon mode to the impurity site:
\begin{equation}
  \label{ah}
  \begin{split}
  H=&\sum_{\sigma}\epsilon_f f^{\dagger}_\sigma f^{}_\sigma 
  + g (b^{\dagger}+b)\sum_{\sigma} (f^{\dagger}_\sigma f^{}_\sigma -\frac{1}{2})\\
  &+\sum_{{\bf k},\sigma}V_{\bf k}(f^{\dagger}_\sigma c^{}_{{\bf k}\sigma}
  + c^{\dagger}_{{\bf k}\sigma}f^{}_\sigma)+\sum_{{\bf k}\sigma}
  \epsilon_{\bf k}c^{\dagger}_{{\bf k}\sigma}c^{}_{{\bf k}\sigma}
  + \omega_0 b^{\dagger}b.
\end{split}
\end{equation}
An extensive study and discussion of this model is presented in
Ref.~\onlinecite{HM02}, which also gives details of the generalization of the
NRG to this situation
\footnote{The main difficulty in extending the NRG to the
model~(\ref{ah}) lies in the unlimited number of bosonic degrees of freedom
already for an isolated impurity. To
perform the NRG, one has to limit the number of bosonic states. As discussed in
Ref.\cite{HM02}, it should
usually be sufficient to limit the number of allowed phonons to $\nu=4 \bar{n}=4
\frac{g^2}{\omega_0^2}$, where $\bar{n}$ is the average number of excited
phonons. However, we usually allow $\nu=200$($\gg\bar{n}$ for all
situations in this paper).}.
In this letter, we present and discuss
results obtained for the (lattice) Holstein
model~(\ref{eq:hamil}) using the NRG in conjunction with the DMFT.
\begin{figure}[t]
  \begin{center}
    \includegraphics[width=0.4\textwidth]{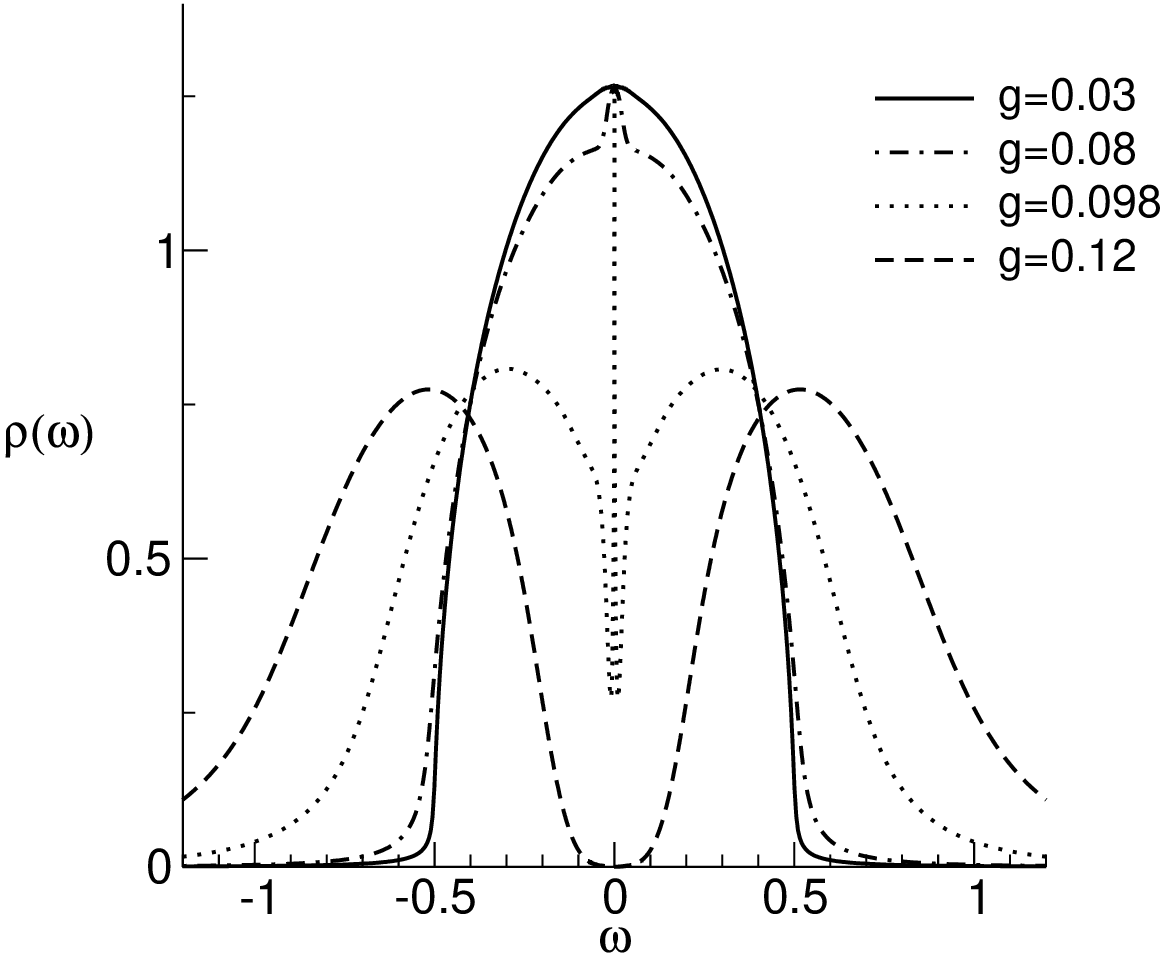}
    \caption{Electronic spectral density $\rho(\omega)=- \frac{1}{\pi}{\rm
        Im}\langle \langle c_{i\sigma}; c_{i\sigma}^\dagger \rangle\rangle$  from the DMFT-NRG calculation for
      $\omega_0=0.05$ and various coupling strengths $g$.}
    \label{fig:dos}
  \end{center}
\end{figure}

We calculate the single-electron spectral function
self-consistently within the DMFT approach. The local phonon
propagator $d(\omega) = \langle\langle b_i,b_i^\dagger \rangle\rangle$ for the Holstein model as
well as spin- and charge susceptibilities for the associated impurity model can
easily be obtained by this method.
The phonon propagator $d(\omega)$ can also be calculated from the charge
susceptibility of the impurity model:
\begin{equation}
  \label{eq:phonprop}
  d(\omega)= d_0(\omega) + g^2 d_0(\omega)^2 \chi_c(\omega),
\end{equation}
where $d_0(\omega)=(\omega-\omega_0+i0^+)^{-1}$ is the phonon propagator for
$g=0$, and $\chi_c(\omega)$ is the charge susceptibility of the associated
impurity model. Another frequently discussed phonon propagator, 
$D(\omega)= \langle\langle (b_i+b_i^\dagger),(b_i+b_i^\dagger) \rangle\rangle$ can be
calculated by a similar formula, which is obtained by replacing $d_0(\omega)$ by 
$D_0(\omega)=2 \omega_0 / (\omega^2-\omega_0^2+i0^+)$ in Eq.~(\ref{eq:phonprop}).

In the following numerical results, we chose for the uncorrelated system ($g=0$)
a semielliptic density of states for
the conduction band. Its width $W=1$ defines the
energy unit used throughout this paper. 
We only consider the particle-hole
symmetric case and, in analogy to the Mott-transition
in the Hubbard model in $d=\infty$, suppress long-range
order, which
corresponds to suppressing anti-ferromagnetic order in the Hubbard model. Unless
otherwise noted, the phonon frequency is taken to be $\omega_0=0.05$.
All calculations were performed for $T=0$, but our
method can be extended to finite temperatures~\cite{BCV01}.

In Fig.~\ref{fig:dos}, the electronic spectral function is plotted for various
values of $g$ with $\omega_0=0.05$. For weak coupling, a small feature emerges at the
Fermi energy ($\omega=0$). With increasing $g$, this peak becomes narrower and
more pronounced. This behaviour is qualitatively similar to that found within
the Migdal-Eliashberg (ME) approach~\cite{HD01pre}, the quantitative difference is
the enhanced narrowing in the NRG calculations.

At intermediate coupling, the central feature becomes very narrow, and two broad
peaks emerge above and below the Fermi energy. These are entirely absent in the
ME approach. At some critical coupling $g_c\approx 0.099$, the central peak
disappears and a gap opens between the two upper and lower bands
\footnote{Note that only for $d=\infty$ this corresponds to a metal-insulator
  transition since here pair hopping is suppressed.}.
For all $g<g_c$, the system is a Fermi liquid with 
${\rm Im} \Sigma(\omega)\sim \omega^2$
for small energies. This manifests itself in Fig.~\ref{fig:dos} in the pinning
of the spectral function at the Fermi energy.
\begin{figure}[t]
  \begin{center}
    \includegraphics[width=0.4\textwidth]{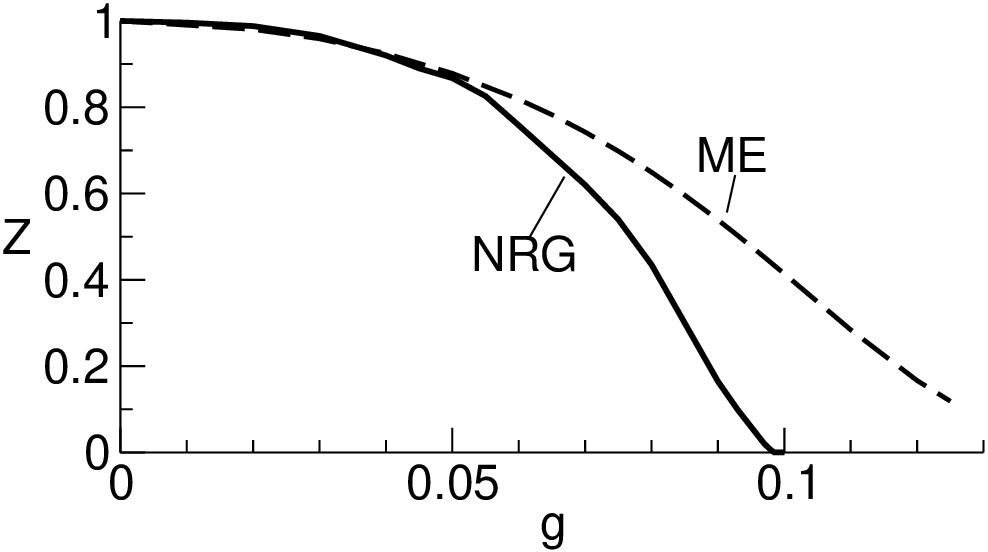}
    \caption{Quasiparticle weight $Z=(1-\frac{\partial \Sigma(\omega)}{\partial
        \omega}|_{\omega=0})^{-1}$ as function of $g$ from the NRG and the
      Migdal-Eliashberg (ME) calculation}
    \label{fig:qweight}
  \end{center}
\end{figure}

The quasiparticle weight $Z=(1-\frac{\partial \Sigma(\omega)}{\partial
  \omega}|_{\omega=0})^{-1}$ is shown in Fig.~\ref{fig:qweight} as obtained
within the NRG and the ME calculation. In both cases, $Z$ decreases with
increasing $g$. Up to $g\approx 0.05$ both lines coincide, but then the NRG
curve decreases faster. The ME calculation breaks down before $Z$ reaches $0.1$.

\begin{figure}[t]
  \begin{center}
    \includegraphics[width=0.4\textwidth]{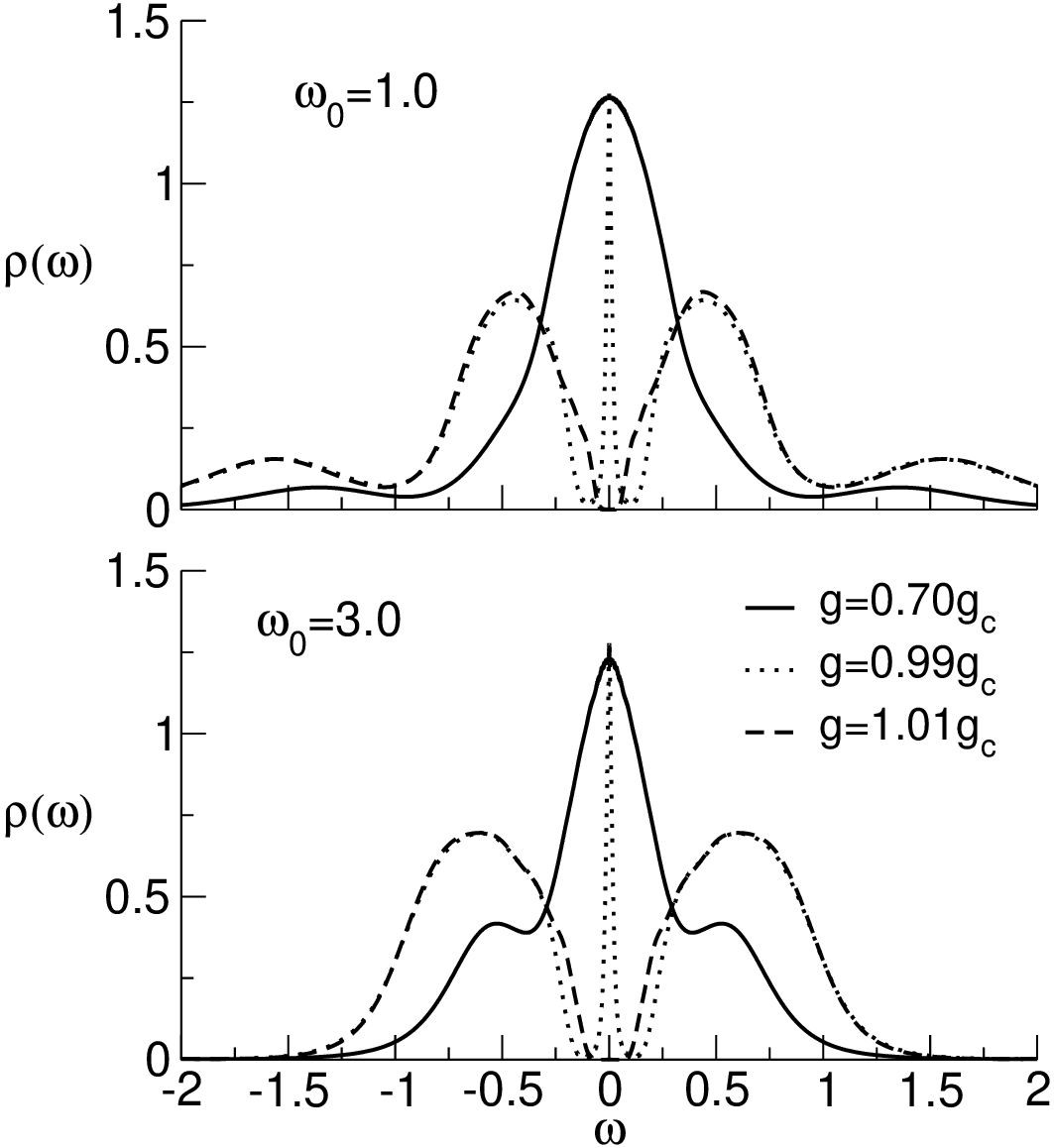}
    \caption{Electronic spectral function as in Fig.~\ref{fig:dos}, but for
      larger $\omega_0=1$ and $3$. The values of $g$ are given in units of $g_c$
      which is $g_c=0.69$ ($1.48$) for $\omega_0=1$ ($3$).}
    \label{fig:dos2}
  \end{center}
\end{figure}
We take a closer look at the large $\omega_0$ limit in Fig.~\ref{fig:dos2},
where the electronic spectral function for $\omega_0=1$ and $3$ are plotted for
three values of $g$ well below, close to, and above $g_c$. The two broad bands
discussed above for $\omega_0=0.05$ split now into two each. For
$\omega_0=3$, the higher-energy ones are not visible any more on the plotted
scale. These bands are \textit{multiphonon bands} and are shifted approximately by
$\omega_0$.
The lower-energy ones are \textit{bipolaron bands}
located at the \textit{bipolaron binding energy} 
$\lambda=2 g^2/\omega_0$. The weight of the multiphonon bands vanishes
with increasing $\omega_0$~\cite{HM02}, and for sufficiently large $\omega_0$,
they can be
neglected. The remaining excitation spectrum corresponds to that of an
attractive Hubbard band where the effective interaction $|U|=\lambda$
corresponds to the
bipolaron binding energy.
 
The Mott metal-insulator transition in the Hubbard model shows
a parameter regime with $U_{c1}<U<U_c=U_{c2}$ where metallic and insulating
solutions co-exist ('hysteresis').
For large $\omega_0$, the effective
$|U_{c1,2}|=2g_{c1,2}^2/\omega_0$ of the Holstein model coincide with the values
known from the Hubbard
model~\cite{Bul99}. For smaller $\omega_0$, the hysteresis region shrinks, and
finally, for $\omega_0=0.05$ no hysteresis is detectable ($g_{c1}=g_{c2}$).

\begin{figure}[t]
  \begin{center}
    \includegraphics[width=0.4\textwidth]{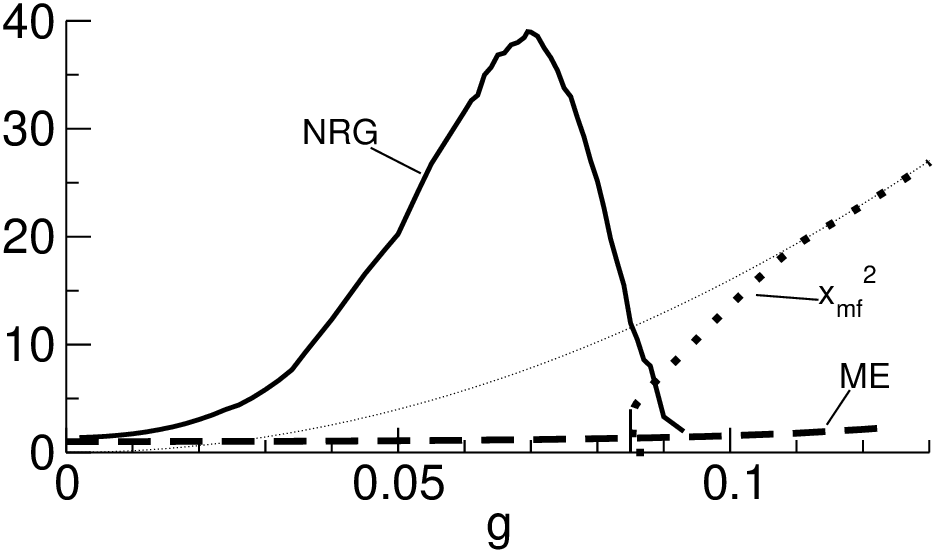}
    \caption{Lattice fluctuations $\langle
      \hat{x}^2-\langle\hat{x}\rangle^2\rangle$ for NRG and Migdal-Eliashberg
      (ME) calculation. Additionally the value of $ x^2_{\rm mf}$ as
      obtained within the mean-field calculation is plotted using a thick dotted
      line. The thin dotted line shows the limiting behaviour of $x^2_{\rm mf}$ for large $g$, $2\frac{g^2}{\omega_0^3}$.}
    \label{fig:xsq}
  \end{center}
\end{figure}

To gain further insight into the results presented so far, let us look at the
mean-field solution of model~(\ref{eq:hamil}).
The classical field $x$ is
self-consistently determined as 
$x = \frac{1}{\sqrt{2 \omega_0}}\langle b + b^\dagger \rangle= -
\sqrt{\frac{2}{\omega_0}}\frac{g}{\omega_0} \langle \sum_\sigma (n_\sigma - \frac{1}{2})\rangle$.
For small $g$ the excitation spectra remain
unchanged from the $g=0$ case since any effects of the distortion are cancelled
out by a change in the chemical potential. The system becomes unstable towards
charge-order at a
critical coupling $g_c^{\rm (mf)}$ which for $\omega_0=0.05$ is $g_c^{\rm
  (mf)}=0.085$. If one (artificially) restores the symmetry, one obtains an electronic
excitation spectrum consisting of 2 peaks shifted by $g x_{\rm mf}$ above and below the
Fermi energy. The self-consistently calculated values of $x_{\rm mf}^2$ are shown in
Fig.~\ref{fig:xsq} together with the asymptotic behaviour for large $g$,
$x_{\rm mf}^2\rightarrow 2 g^2/\omega_0^3$.

This behaviour can be explained by thinking in terms of an effective potential
for the ions $V(x)$.
For $g<g_c^{\rm (mf)}$, the $V(x)$ is a simple harmonic
potential. For 
$g>g_c^{\rm (mf)}$, it changes into a double-well structure with
minima at $x_{1,2}=\pm g x_{\rm mf}$. In the mean-field approximation
no fluctuations between these minima occur.
To go beyond this, one needs to include these lattice fluctuations between the two
minima, which has been considered in 
Ref.~\onlinecite{BZ98}.

The magnitude of the lattice fluctuations, $\langle \hat{x}^2 - \langle \hat{x}
\rangle ^2\rangle$  is plotted in Fig.~\ref{fig:xsq}. Within the NRG
calculation, this quantity has a clear maximum at a  value $g^*$ ($<g_c$).
At $g_c$, the fluctuations are already
significantly reduced. 
The maximum occurs in the crossover region from a single- to the double-well
potential, where the effective potential is broad and shallow.
The potential barrier then grows rapidly with increasing $g$. 
The corresponding fluctuations in the ME calculation are always small.

\begin{figure}[t]
  \begin{center}
    \includegraphics[width=0.4\textwidth]{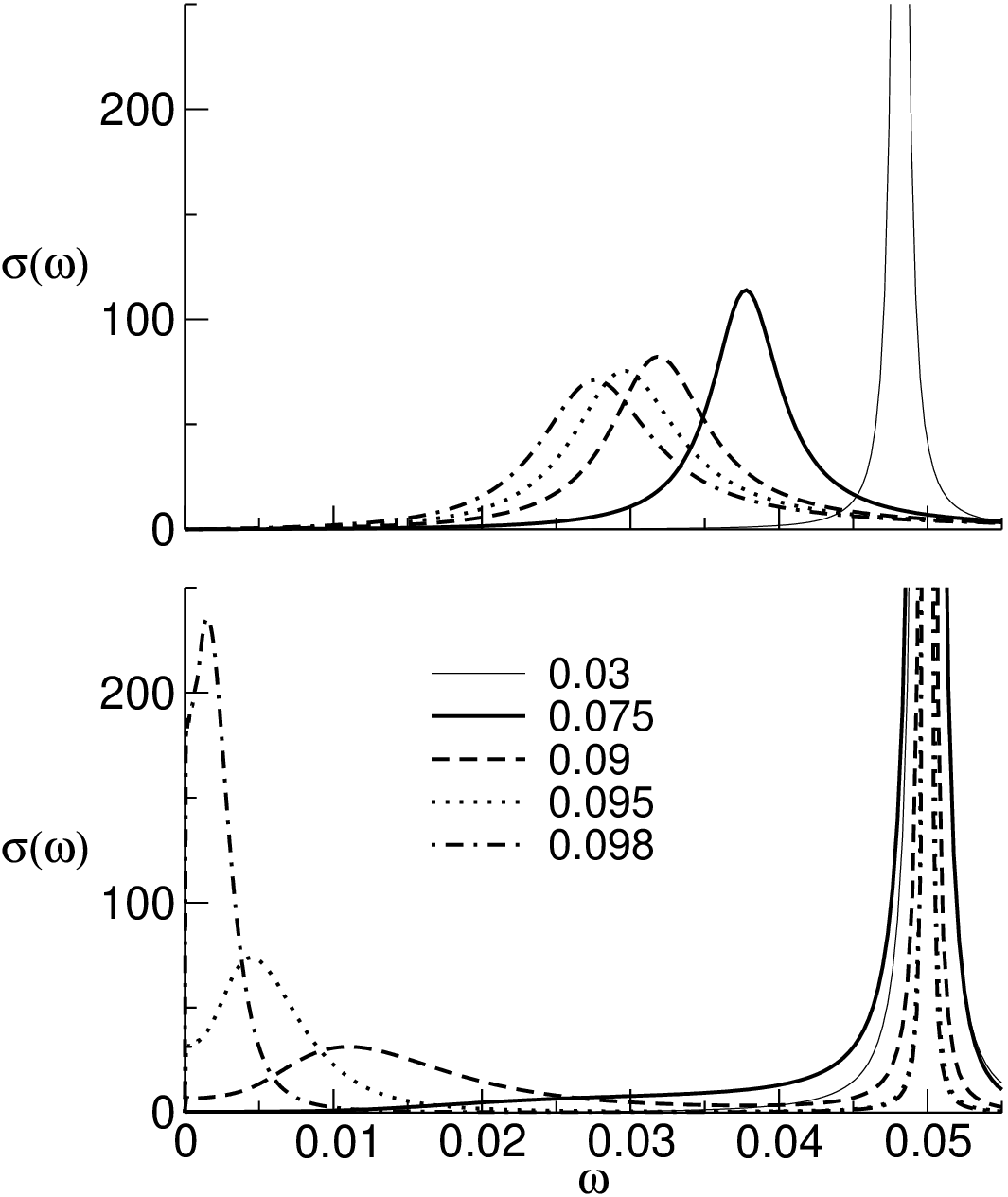}
    \caption{Phonon propagator $\sigma(\omega)=-\frac{1}{\pi}{\rm Im} \,d(\omega)$ for $\omega_0=0.05$ and various coupling strengths
      $g$. The upper panel shows the Migdal-Eliashberg, the lower panel the NRG results.}
    \label{fig:phonprop}
  \end{center}
\end{figure}
Figure~\ref{fig:phonprop} shows the
phonon propagator  ${\rm Im} \, d(\omega)$ both for the Migdal-Eliashberg and the
DMFT-NRG calculation. The two methods give completely different pictures. In the
ME approach, the phonon propagator, which for $g=0$ consists of a peak at
$\omega=\omega_0$, remains a single peak which softens with increasing coupling
strength. In contrast to
that, the NRG result shows that the main peak at $\omega=\omega_0$ broadens, but
remains essentially unshifted. In addition, a second phonon mode appears at
lower energy with increasing coupling strength. As $g$ approaches $g_c$ this
mode softens and diverges at
$g=g_c$. In the insulating phase, only the peak at $\omega=\omega_0$ remains and
narrows. This behaviour in the insulating phase is to be expected as the opening
of the gap inhibits any broadening due to electron-hole excitations.

Closer inspection of Figs.~\ref{fig:xsq} and \ref{fig:phonprop} shows that the
soft phonon mode develops for $g\approx g^*$, corresponding to the maximum in 
$\langle \hat{x}^2-\langle\hat{x}\rangle^2\rangle$. This soft mode thus
coincides with the build-up
of the potential barrier in the effective potential of the phonons.
From Eq.~(\ref{eq:phonprop}) it follows directly that the phonon
propagator is closely related to the charge susceptibility of the associated
impurity model. The peak in ${\rm Im} \,d(\omega)$ has its equivalence in the
low-energy peak of $\chi_c$. The existence of such a low-energy peak
can be expected as follows from the mapping of the Holstein onto
an attractive 
Hubbard model. As noted before, the physics of the attractive Hubbard model
correspond exactly to those of a repulsive Hubbard model with the spin- and
charge-channels exchanged. And the latter should have a low-energy peak in the spin
susceptibility due to the Kondo physics of its associated impurity model.
The physics of the gap formation, and its precursor regime in the Holstein model
are thus dominated by many-body physics. The low-energy feature in the phonon
propagator has not been predicted before.

In this letter we have presented the application of the dynamical mean-field theory in
combination with the numerical renormalization group to the Holstein model at
half-filling. This
method can be applied to essentially all parameter regions of
the model. 
We studied the gap formation for small and large phonon frequencies.
Generally, the gap formation has precursor effects due to
many-body 'Kondo-like' physics: very strong lattice fluctuations indicate
formation of a double-well potential for the ions, and a soft phonon mode
emerges due to fluctuations between the two states of the system.
It might be possible to observe it experimentally.
The framework of our method can be extended to contain other local
interactions such as electron-electron interaction of Hubbard-type to describe
fullerides, or spin-exchange interactions as used to describe the manganites.
\begin{acknowledgments}
We wish to thank the EPSRC (Grant GR/J85349) and the DFG (SFB 484) for financial
support, and
D. M. Edwards and J. Freericks for stimulating discussions.
\end{acknowledgments}

\end{document}